# WASP-12b: A Mass-Losing Extremely Hot Jupiter

Carole A. Haswell
School of Physical Sciences, The Open University, Walton Hall, Milton Keynes, MK7 6AA, United Kingdom
e-mail: Carole.Haswell@open.ac.uk

**Abstract**

WASP-12b is an extreme hot Jupiter in a 1 day orbit, suffering profound irradiation from its F type host star. The planet is surrounded by a translucent exosphere which overfills the Roche lobe and produces line-blanketing absorption in the near-UV. The planet is losing mass. Another unusual property of the WASP-12 system is that observed chromospheric emission from the star is anomalously low: WASP-12 is an extreme outlier amongst thousands of stars when the log $R'_{HK}$ chromospheric activity indicator is considered. Occam's razor suggests these two extremely rare properties coincide in this system because they are causally related. The absence of the expected chromospheric emission is attributable to absorption by a diffuse circumstellar gas shroud which surrounds the entire planetary system and fills our line of sight to the chromospherically active regions of the star. This circumstellar gas shroud is probably fed by mass loss from WASP-12b. The orbital eccentricity of WASP-12b is small but may be non-zero. The planet is part of a hierarchical quadruple system; its current orbit is consistent with prior secular dynamical evolution leading to a highly eccentric orbit followed by tidal circularization. When compared with the Galaxy's population of planets, WASP-12b lies on the upper boundary of the sub-Jovian desert in both the ($M_P$, $P$) and ($R_P$, $P$) planes. Determining the mass loss rate for WASP-12b will illuminate the mechanism(s) responsible for the sub-Jovian desert.

## Introduction

In 1995 our preconceptions about planetary systems were challenged with the discovery of 51 Peg b, the first hot Jupiter (Mayor and Queloz 1995). The subject of this Chapter, WASP-12b, is one of the hottest of the hot Jupiters, and its extreme properties have encouraged many studies. This contribution focuses on the phenomenon of mass loss from this extreme planet and on its implications for the exoplanet population, without attempting to be an exhaustive review of other properties. WASP-12b was an early discovery by the SuperWASP planet search; its large radius, $R_P = 1.7\ R_J$, and short orbital period, $P=1.09$ d, both contribute to favourable probabilities for detection by transit (see e.g. Haswell 2010 for details.) WASP-12b orbits an F star (Hebb et al 2009), which, along with the very short orbital period, means it is among the most irradiated of the known exoplanets. The planet mass is $M_P = 1.4\ M_J$ and the orbital semi-major axis is 0.023 AU. Of these planet parameters, as so often in astronomy, the period is by far the best determined. The other planet parameters rest upon our knowledge of the basic stellar parameters, and their reported uncertainties often do not completely propagate our uncertainties about the host star. Haswell (2010) gives an explanation of the derivation of the fundamental parameters for transiting planets.

The host star of this extreme hot Jupiter, WASP-12, has effective temperature $T_{eff}$ = 6250±100 K and surface gravity log $g$ = 4.2± 0.2 and is metal rich compared to the Sun; the star's age is between 1 and 2.65 Gyr and its mass $M_*$ is between 1.23$M_\odot$ and 1.49 $M_\odot$ (Fossati et al 2010a, Hebb et al 2009). The planet's orbit places it only about 1.5 stellar radii above the star's photosphere, making it an extreme hot Jupiter. Swain et al (2013) found that WASP-12b's near-IR brightness temperature is around 3000K, approximately twice that of the archetypal hot Jupiters HD 189733b and HD 209458b. The distance to WASP-12 is less precisely known, with Fossati et al (2010a) finding a possible range of 295-465 pc. These limits are based on the star's securely measured effective temperature and surface gravity; narrower uncertainty ranges have been quoted elsewhere, but these ranges may rest upon debatable assumptions. See Fossati et al (2010a) for a detailed discussion of WASP-12's fundamental parameters, including the distance.

The seminal UV observations of the first known transiting exoplanet, HD 209458b, by Vidal-Madjar et al (2003), showed that hot Jupiter planets are surrounded by extensive exospheres. HD 209458b produces a 15% deep transit in the Lyman α line, indicating the planet is surrounded by an exosphere with radius > 3 $R_P$ and containing neutral hydrogen. UV observations underpin the study of exoplanet exospheres because this spectral region contains strong resonance lines of many common species. Strong resonance lines have large oscillator strengths and involve the ground state and can thus be detected even when the emitting or absorbing column density is modest. Consequently, these lines are (i) prominent in the emission from stellar chromospheres and (ii) provide a particularly sensitive probe for the presence of diffuse gas surrounding a planet. Taken together, this means the UV contains many features which are particularly useful for transmission spectroscopy of transiting planets.

## UV Observations of WASP-12b transits

It was immediately clear upon the discovery of WASP-12b that this system is a good candidate to search for exospheric gas surrounding the planet. The short orbital period combined with the inflated planet radius mean that the planet fills a substantial fraction of its Roche lobe. WASP-12b remains one of the three known giant planets with the largest Roche lobe filling factor (Busuttil 2017). Hubble Space Telescope (HST) observations were promptly proposed after the discovery. The obvious goal was to perform transmission spectroscopy covering Lyman α to determine the extent of the HI exosphere for such a strongly irradiated planet. This proved impractical because WASP-12 is 6 to 10 times more distant than HD 209458b and consequently the predicted stellar photon count-rate for WASP-12 was too low. HD 209458b at 47 pc is close enough for a precise distance determination by Hipparcos parallax measurements, so produced good far UV count-rates despite there being so little stellar flux in the far-UV region which contains Lyman α. To overcome the distance of WASP-12 and the intrinsic faintness of FGK stars in the far UV, the first HST observations of WASP-12 pioneered HST transmission spectroscopy in the near-UV wavelength range, λλ 2539 – 2829 Å (Haswell et al 2012) where the underlying stellar photospheric flux is enhanced by a factor of ~$10^6$. This proved to be an extremely informative wavelength region. Despite this, it has remained under-exploited in subsequent studies of exoplanet exospheres.

*Hubble Space Telescope Cosmic Origins Spectrograph observations of WASP-12b*

The observations presented in Haswell et al (2012) are a superset of those presented in Fossati et al (2010b) and used the slitless Cosmic Origins Spectrograph (COS) to obtain 10 HST orbits of R~20,000 near-UV spectroscopy. COS records data in three wavelength regions known as 'stripes' A, B and C. This data was obtained on two distinct HST visits, each centred on a transit of WASP-12b; the visits were timed so that the orbital coverage of WASP-12b interleaves. The two visits used slightly different grating settings, so that each of the 3 stripes only partially overlaps in wavelength between the two visits. The significantly noiser data subsequently obtained and reported by Nichols et al (2015) used exclusively the Visit 1 setting of Haswell et al (2012).

The near-UV spectral region contains thousands of overlapping photospheric absorption lines. The strongest of these are resonance lines, including the Mg II h&k lines, Mn II 2577Å and Fe II 2586Å which can be easily picked out in Figure 17 of Haswell et al (2012). Unevolved stars generally exhibit chromospheric emission in the line cores of these features. The plethora of weaker lines blend to produce substantial stellar photospheric absorption of the stellar continuum which illuminates the bottom of the stellar photosphere. This line-blanketing stellar photospheric absorption is strongest in the C stripe, and weakest but still approaching 50% in the B stripe, as shown in Figure 1 of Haswell et al (2012).

*Photometric Analysis*

Because slightly different wavelength regions were covered, it is impossible to produce a set of homogeneous light curves including all the photons collected in the observations reported in Haswell et al (2012). The comprehensive analysis is described in detail in the original paper, the broad conclusions are summarised here.

In both visits, the transit of WASP-12b in the near-UV region is deeper than that of the opaque planet, indicating that the planet is surrounded by an exosphere of strongly absorbing gas. In the first visit, reported by Fossati et al (2010b), the relative depths of the transits in the A, B, and C stripes corresponds roughly to the relative amounts of absorption in these regions within the stellar atmosphere. The near-UV-absorbing gas thus appears to have a temperature and composition similar to those of a stellar photosphere. This finding was independently corroborated by the interpretation of IR transmission spectroscopy and secondary eclipse data by Swain et al (2013). The transit depth in the A and C stripes of Visit 1 of Haswell et al (2012) is 3.5%, which makes it clear that the absorbing gas surrounding WASP-12b overfills the planet's Roche lobe, at least some of the time. In turn, this implies that WASP-12b is losing mass.

In detail, the repeated observations reported in Haswell et al (2012) and Nichols et al (2015) do not reproduce the shape of the Visit 1 transit light curve. In particular the early ingress for Visit 1 reported in Fossati et al (2010b) is not a general feature. Visit 2 of Haswell et al (2012) instead shows a high point at the phase of optical ingress: Visit 2 and the four visits of Nichols et al (2015) suggest the absorbing gas is patchy, and its spatial coverage as seen from our vantage point extends to the earliest observed orbital phases, at around $\varphi=0.83$. Figure 9 of Carroll-Nellenback et al (2017) visualises the dispersed material lost in a hydrodynamic outflow from a hot Jupiter. It shows a patchy distribution of material which extends azimuthally to surround the star. The patchy and variable absorption observed in the near-UV lightcurves of WASP-12, as described in Haswell et al (2012) and Nichols et al (2015), is consistent with this picture.

*Spectroscopic Analysis*

Despite the extension of absorbing gas spread azimuthally around the orbit of WASP-12b, there is a consistent relative overdensity around the planet itself. There is clear evidence for this: the near-UV light curves consistently show fluxes during transit which are depressed by more than the 1.5% dip caused by the opaque planet itself (Fossati et al 2010b, Haswell et al 2012, Nichols et al 2015). This means that the unocculted light from the star suffers from increased absorption at phases where the planet is in transit compared to that at other observed phases. This can be seen for individual lines, too: there is evidence from the very strong Mg II h&k lines that there is some absorption from diffuse gas at all observed phases (see below), but despite this we can perform transmission spectroscopy by comparing spectra obtained during transit with those obtained away from transit. The comparison allows us to identify excess absorption attributable to the higher column density of intervening diffuse gas during the planet transit.

Fossati et al (2010b) performed transmission spectroscopy of WASP-12b for Visit 1 of Haswell et al (2012) finding enhanced transit depths at the wavelengths of resonance lines of neutral sodium, tin, and manganese, and of singly ionised ytterbium, scandium, manganese, aluminium, vanadium and magnesium. Haswell et al (2012) detected repeated enhanced absorption during transit in 65 distinct features. These features include exospheric absorption throughout the inner wings of the very strong Mg II h&k lines on both visits and a detection of exospheric absorption in Fe II 2586Å, which remains the heaviest species detected in an exoplanet transit. Absorption at these individual features is accompanied by line-blanketing absorption which accumulates to produce the broad-band absorption seen in the near-UV light curves. Not all the features in the WASP-12b transmission spectrum can be unambiguously identified (Haswell et al 2012). In particular, in some cases it is unclear whether a feature is due to a resonance line of a rare element or a weaker feature of a more abundant element, or a blend of the two. These cases should be revisited in the future when we have a better understanding of the composition of WASP-12b's outer layers and the mechanisms underlying the mass loss which feeds the exospheric gas.

*WASP-12's missing chromospheric emission in MgII h&k*

All the observations of WASP-12's Mg II h&k lines show zero flux in the line cores (Fossati et al 2010b, Haswell et al 2012, Nichols et al 2015). The strong Fe II 2586Å line, covered only in Visit 2 of Haswell et al 2012, is also consistent with zero flux in the line core. This makes WASP-12's observed spectrum unique: all other main sequence or slightly evolved stars of WASP-12's spectral type show chromospheric emission in these features. A basal level of chromospheric flux is seen even for old, slowly rotating, inactive stars of this type.

Occam's razor suggests that this anomaly in WASP-12's spectrum is related to the extreme exoplanet it hosts. Haswell et al (2012) examined the possibility that the Mg II h&k line core flux is absorbed by the interstellar medium (ISM), concluding that an anomalously over-dense line of sight would be required. The missing chromospheric emission in Mg II h&k and the Fe II 2586Å line is more likely to be absorbed locally by a diffuse circumstellar gas shroud fed by the mass-loss from the overflowing exosphere of the extreme hot Jupiter planet WASP-12b.

## Stellar Chromospheric emission shrouded by planetary mass loss

The Mg II h&k lines can only be observed from space because their wavelengths are in the UV. However, the analogous resonance lines of the singly ionised element from the next period in the periodic table, Ca, are in the optical and are accessible from the ground. Consequently, the Ca II H&K lines ($\lambda\lambda$ 3968Å, 3933Å) are commonly-used indicators of stellar chromospheric activity, usually parameterised in terms of log $R'_{HK}$ (Noyes et al 1984). This particular metric is useful because it allows intercomparison of the activity levels of F, G, and K type stars on a conveniently calibrated scale; see Staab et al (2017), their Figures 2 & 7, for an illustration of how log $R'_{HK}$ is defined. In particular, for a main sequence FGK star, log $R'_{HK}$ is expected to exceed a value of -5.1 which corresponds to a basal level of activity observed in the oldest, least active, slowest rotating stars of this type. The basal emission persists even when stars are completely devoid of active regions (Schröder et al 2012). WASP-12 lies significantly below this limit with log $R'_{HK}$= -5.5, a value determined by Knutson et al (2010) who did not remark upon its anomalous nature.

There has been significant interest in the measured log $R'_{HK}$ values of transiting planet host stars as there appears to be a correlation between the host star's log $R'_{HK}$ value and the planet's surface gravity (Hartman 2010, Figueira et al 2014, Fossati, Ingrassia & Lanza 2015). Fossati et al (2013) and Staab et al (2017) compared log $R'_{HK}$ for WASP-12 and other planet hosts with an extensive sample of field stars. While Staab et al (2017) show that a quarter of known host stars of close-in transiting planets show anomalously low Ca II H&K chromospheric emission, WASP-12 remains the most extreme outlier.

Fossati et al (2013) and Staab et al (2017) interpret the missing Ca II H&K chromospheric emission in these close-in planet host stars as due to absorption by diffuse gas shrouding these planetary systems. This builds on and corroborates the conclusions Haswell et al (2012) drew from WASP-12's missing Mg II h&k and Fe II chromospheric emission. In each of these low log $R'_{HK}$ systems, gas originates in an outflow from the close-in ablating planet, forming a diffuse cloud of circumstellar gas which shrouds the entire planetary system. It is plausible that the extremely irradiated planet WASP-12b has a more prodigious mass outflow than more moderately irradiated giant planets. While the detailed dynamics of these outflows is largely unexplored, it is natural to expect this to lead to a higher column density of gas in the line of sight to the chromospherically active regions of this particular star.

Significantly, Staab et al (2017) show that two low-mass close-in planet hosts, Kepler-25 and Kepler-28, have anomalously low values of log $R'_{HK}$, establishing that the phenomenon does not require the presence of a gas giant planet. Indeed the most observationally spectacular examples of ablating planets are the catastrophically disintegrating exoplanets (CDEs), exemplified by Kepler 1520b. The CDEs are extremely close-in planets, detected via their periodic but variable transits. The transits are attributed to a cloud of dust formed by material ablated from the rocky surface of a low mass planet irradiated to temperature $T$~2100K. In the case of Kepler 1520b, the planet is thought to have a mass of about $0.1 M_\oplus$. At temperatures of ~2100K rocky minerals sublime, and a thermal wind of metal-rich vapour and entrained dust flows from the planet. The entrained dust can vary in spatial distribution and optical depth, thus producing variable transits (Rappaport et al 2012). The transiting dust cloud model was confirmed when Bochinski et al (2015) detected the colour dependence of transit depth of Kepler 1520b. The known CDE host stars are sadly too distant and faint to encourage transmission spectroscopy with current facilities.

## WASP-12b and Mass Loss in Hot Jupiter Planets

Since the discovery of the extended HI exosphere of HD 209458b there has been significant interest in the phenomenon of mass loss in hot Jupiter planets. In this section we will discuss first the models, then the relationship between the models and WASP-12b, which is the most extreme known hot Jupiter, and finally we will consider the potential future observations of WASP-12b which might clarify some of the presently open questions.

### *Mass-loss rates in hot Jupiters: brief overview of the state of the art*

The discovery of the extended HI exosphere of HD 209458b stimulated prompt modelling work to calculate atmospheric loss from hot Jupiters (e.g. Lammer et al 2003). There has been much activity in this area in the intervening years, with recent papers including Jackson et al (2017) and Carroll-Nellenback et al (2017). Lammer et al (2003) showed that irradiation by stellar X-ray and extreme-UV (EUV) flux leads to atmospheric expansion, and mass-loss rates of ≈$10^{12}$ g s$^{-1}$ for HD 209458b. Later work (e.g. Erkaev et al 2007) incorporated the Roche potential, which lowers the energy required for atmospheric escape. In the extreme case of a Roche lobe-filling planet, there is no energy barrier to atmospheric escape through the saddle point at $L_1$. Recent models including Tripathi et al (2015) and Jackson et al (2017) variously incorporate Roche lobe overflow, the evolutionary response of the planet to mass loss, the irradiative heating and the three-dimensional hydrodynamic outflow. Computational limitations mean that no single model has yet simultaneously included all of these ingredients self-consistently, but the crucial factors which determine the mass-loss rate for close-in planets are being established and quantitatively tested.

As laid out in Frank et al (2002), the mass loss rate depends critically on the comparision between the evolution of the mass donor's size and the evolution of the Roche lobe size. This in turn depends on the transfer of orbital angular momentum within the system. Valsecchi et al (2015) investigated tidally driven Roche lobe overflow of hot Jupiters within this framework, including the effects of irradiation and planet evolution. They produce evolutionary tracks tracing the mass, radius, and orbital period for 1 $M_J$ planets orbiting 1$M_\odot$ stars, both for conservative and non-conservative mass transfer. In conservative mass transfer all mass leaving the planet is accreted onto the star; conversely, non-conservative mass transfer means some fraction of the mass from the planet leaves the system. In non-conservative mass transfer, angular momentum will generally be lost.

Observational evidence in WASP-12 and other irradiated planet host stars for the presence of circumstellar gas shrouds (Haswell et al 2012, Fossati et al 2013, Staab et al 2017) implies that mass transfer is non-conservative. This has implications for the planets' orbital evolution, which transit timing measurements over long temporal baselines can test.

*The case of WASP-12b*

As noted before, WASP-12b is among the most irradiated of exoplanets; temperatures in the outer layers of the planet's atmosphere are in the same regime as M dwarf star photospheres. It is an extremely hot Jupiter. As shown in Figure 21 of Haswell et al (2012) WASP-12b is close to filling its Roche lobe, with a volume filling fraction of 0.54 (Busuttil 2017). A planet in this configuration is expected to lose mass through hydrodynamic escape. Stellar irradiation will photoionize the upper atmosphere, with the liberated electrons causing collisional heating. This heating drives a hydrodynamic wind comprised of the constituents of the upper atmospheric gas (c.f. Erkaev et al 2007, Sanz-Forcada et al 2010, Tripathi et al 2015; Jackson et al 2017.) Bisikalo et al (2013) performed a hydrodynamic simulation of outflows from both the $L_1$ and $L_2$ points of WASP-12b's Roche lobe, concluding that a steady-state gaseous envelope encompassing the Roche lobe can result. This simulation did not include treatment of the photoionisation which drives the wind, instead setting *a priori* a temperature of $10^4$K outside the Roche lobe. More recent work by Carroll-Nellenback et al (2017) has taken a more self-consistent approach to the dynamics of similar 'up-orbit' and 'down-orbit' outflows, though not tailored to the specific case of WASP-12b.

Husnoo et al (2011), Albrecht et al (2012) and Collins et al (2017) analysed radial velocity data of WASP-12 from SOPHIE in which the Rossiter McLaughlin effect is of comparable amplitude to the systematic instrumental effects. Thus the spin of WASP-12 appears to be slow or approximately orthogonal to the orbital angular momentum of WASP-12b's orbit. For a slowly-spinning host star under the assumptions of Valsecchi et al (2015), tides transfer angular momentum from the orbit to the stellar spin, causing orbital decay. For WASP-12b, there is evidence that the orbit may be tidally shrinking in this way (Maciejewski et al 2016), but see also Collins et al (2017).

All these considerations imply that WASP-12b is likely to be shedding mass more rapidly than (almost?) all other known exoplanets. This is consistent with the detection of gas surrounding the planet during transit, which presents an effective obscuring area in the near-UV of up to three times that of the optically opaque planet (Haswell et al 2012), and in excess of the cross-section presented by the planet's Roche lobe. The anomalously depressed stellar chromospheric emission from WASP-12 at all observed orbital phases is also consistent with dispersed gas shrouding the entire planetary system. Haswell et al (2012), Fossati et al (2013) and Staab et al (2017) show that WASP-12 has the lowest chromospheric emission of any star measured, which suggests that either the intrinsic stellar activity is extremely low, WASP-12b's mass loss is prodigious and produces a higher column depth of shrouding circumstellar gas than is present in any other system, or both. WASP-12b may be in the early stages of the evolutionary phase discussed by Valsecchi et al (2014). They suggest Roche lobe-filling hot Jupiters may lose their entire gaseous envelopes, leaving hot super-Earths.

The 'up-orbit' stream of Carroll-Nellenback et al (2017) can be accelerated away from the star due to ram pressure from the stellar wind. The simulations neglect radiation pressure: the absorption of stellar emission found by Fossati et al (2010b), Haswell et al (2012) and Nichols et al (2015) guarantees radiation pressure will tend to drive WASP-12's circumstellar gas outwards.

The quantitative self-consistent numerical calculations of mass loss rates for particular hot Jupiter systems have generally been tailored to more moderately irradiated planets at larger orbital distances. Bisikalo et al (2013) did not calculate a mass loss rate for WASP-12b. The value of mass loss rate of $10^{-7}$ $M_J$ year$^{-1}$ calculated by Li et al (2010) for WASP-12b was predicated on an erroneous non-zero eccentricity inferred from the discovery paper radial velocity data (Hebb et al 2009). Husnoo et al (2011) subsequently derived e=0.017 $^{+0.015}_{-0.010}$, i.e. an approximately circular orbit. Lai et al (2010) obtained a mass loss rate 25 times lower with assumptions of an isothermal planet atmosphere of 3000K, a density at $L_1$ of $3 \times 10^{12}$ g cm$^{-3}$ and a nozzle for Roche lobe overflow which is of linear size of 0.6 planet radii. It is fair to say not all of these assumptions are supported in detail. For example, using atmospheric structure models motivated by the observations of Swain et al (2013) and Mandell et al (2013) can change the mass loss rate by four orders of magnitude. Thus the mass loss rate for WASP-12b remains undetermined.

*Observations to determine the mass loss rate of WASP-12b*

The flux of ionising stellar radiation upon the planet is a crucial parameter for calculations of the photoevaporative mass loss from hot Jupiters. For stars of WASP-12's spectral type and late type stars generally, this flux largely arises from stellar activity. The Mg II and Ca II lines indicate that WASP-12's apparent chromospheric activity is anomalously low, but unless the star is unique, this must be due to absorption of the intrinsic chromospheric emission. The ionising flux could in principle be directly measured, but WASP-12 is too distant for this to be practical, because ionising flux is strongly absorbed by the ISM (see Fossati et al 2013 for a general discussion of the effects of interstellar absorption on the observed chromospheric emission from exoplanet host stars). An alternative approach would be to take a spectrum in the far-UV (i.e around 1100- 2500 Å) as Fossati et al (2015) have done for WASP-13. This wavelength region contains strong chromospheric emission lines of ionised species including C IV and Si IV. Because these species are rare in the ISM, these lines are not absorbed even for moderate hydrogen column densities. Linsky et al (2013, 2014) observed nearby stars and constructed empirical relationships between the various components of stellar chromospheric emission. Using this, one can take the C IV and Si IV line fluxes and infer the intrinsic fluxes of the remainder of the chromospheric emission. Thus for WASP-13, Fossati et al (2015) were able to derive the intrinsic stellar EUV flux, and subsequently model the upper atmosphere of WASP-13b, deriving a mass-loss rate of $1.5 \times 10^{11}$ g s$^{-1}$.

A good signal to noise spectrum of WASP-12 in the 1100- 2500 Å wavelength region will require a substantial investment of time on HST or a successor UV space telescope. Given the prodigious recent and ongoing modelling activity, this could be a proportionate and appropriate investment with widespread implications.

Optical light curves measuring the transits of WASP-12b are much easier to obtain. Even relatively small telescopes can determine transit timings to useful precision. Maciejewski et al (2016) and Collins et al (2017) analyse optical transit timings, using 31 (23) transit lightcurves spanning Nov 2012 – Feb 2016 (Nov 2009 – Feb 2015) respectively. Both analyses incorporate additional orbital constraints from radial velocity (RV) measurements and secondary eclipse timings, with the analysis of Maciejewski et al (2016) allowing a finite eccentricity in their fit while Collins et al (2017) fix the eccentricity to zero. Collins et al (2017) conclude that the transit timings are consistent with a linear ephemeris, while Maciejewski et al (2016) present evidence for deviations. Maciejewski et al (2016) fit two models: a quadratic ephemeris consistent with a simple tidal decay, and a 3300d periodic signal attributed to the periastron precession of a slightly eccentric orbit with $e=0.00110\pm 0.00036$. Data from winter 2016/17 should distinguish between these two models. Bonomo et al (2017) used transit timings, secondary eclipse timings, all suitable RVs from the literature published up to 1 January 2016 and 15 unpublished HARPS-N RV points taken before February 2015 to perform an Bayesian determination of WASP-12b's orbital parameters, finding $e < 0.020$ (1σ upper limit). Thus, the published data is consistent with a circular orbit for WASP-12b. Knowledge of the current orbital configuration is a vital ingredient to determine the tidal effects on the planet which in turn strongly affect the internal heating and mass loss rate, as dramatically illustrated by Li et al (2010)'s work which assumed a much larger eccentricity. The present orbital configuration can also potentially rule out entire classes of evolutionary scenarios. This in turn has potentially profound implications for the planet structure resulting from the mass loss history, and consequently on the present evolution of the planet radius in response to mass loss. See Valsecchi et al (2015) for modelling work which encapsulates such considerations.

Further high quality radial velocity measurements covering the phases of the WASP-12b transit should allow the stellar spin to be better constrained. The existing SOPHIE observations and analysis (Husnoo et al 2011, Albrecht et al 2012 and Collins et al 2017) show scattered residuals with deviations from the fit comparable in amplitude to the Rossiter McLaughlin effect itself. The superior precision and stability of HARPS-N should allow for a much more precise determination of the projected stellar spin-orbit angle. This information can be used to perform calculations of the tidal transfer of angular meomentum from the orbit to the stellar spin, using techniques similar to those of Valsecchi et al (2015).

## Testing the Pollution Hypothesis

In the currently accepted interpretation of the evidence we have discussed for the WASP-12 system, some fraction of the mass lost from WASP-12b is dispersed outwards, and is probably carrying angular momentum. In the terminology of Frank et al (2002) and Valsecchi et al (2015) this is an example of non-conservative mass transfer. A simple analysis of mass transfer in the Roche geometry reveals that the minimum energy configuration which satisfies the conservation laws is to move a small amount of mass to very large distances. This mass carries the angular momentum away and allows the bulk of the remaining mass to settle into the deepest part of the Roche potential. If this captures the fate of the gas lost from WASP-12b, then the bulk of the material may have accreted onto the star. Fossati et al (2010b) performed a detailed analysis to examine the element by element photospheric abundances of WASP-12 for signs of accretion of material from the planet. Some hints of this atmospheric pollution were found, but a more detailed differential analysis comparing WASP-12 to stellar twins is required for a definitive answer. This would be analogous to the differential abundance analyses of Melendez et al (2009) on solar twins and Nickson (2015) on 61 Virginis and closely matched comparison stars.

## Implications for exoplanet demographics

The historical collection of biological specimens led to an appreciation of the relationships between various species, and ultimately to our understanding of evolution and the expression of genetic characteristics encoded in DNA. Similarly, the purpose behind this exoplanets fauna is to illuminate the relationships between the Galaxy's planetary systems, using well-studied exemplars to reveal the long-timescale evolutionary processes at work. WASP-12b is a useful exemplar of a giant planet suffering high insolation as a result of a very short orbital period around a hot (F-type) host star. It provides an interesting and informative test-case for the mechanisms leading to one of the most prominent features of the emerging demographic information about the Galaxy's population of planets.

### *The Sub-Jovian Desert*

When only 106 transiting exoplanets were known, Szabó & Kiss (2011) identified a very prominent feature in the distribution of planet radii with orbital period. There is a distinct lack of planets with radii between about 2 $R_\oplus$ and 1$R_J$ for orbital periods shorter than about 2.5 days. This feature has subsequently been called the Neptunian desert or the sub-Jovian desert by various authors. As more transiting planets were discovered the feature persisted, as seen, e.g., in Figure 4 of Mazeh et al (2016) showing the $R_P$ vs $P$ distribution for almost 4500 transiting planets and candidate planets from *Kepler*. With the addition of thousands more objects, there are a few within the sub-Jovian desert in the ($R_P$, $P$) plane. In contrast the ($M_P$, $P$) plane shown in Figure 1 of Mazeh et al (2016) includes only the best-studied 1037 transiting planets. In this Figure the sub-Jovian desert is remarkably well-defined.

The sub-Jovian desert cannot be attributed to selection biases: short period planets are favoured by both the radial velocity and the transit detection methods. The former favours massive planets, while the latter favours large planets, and with both methods planets of low mass and / or small radius have been detected below the desert. See, e.g., Haswell (2010) for a discussion of these selection biases.

Mazeh et al (2016) perform a statistical analysis of subsets of the populations they consider, allowing them to derive algebraic forms for the sharply defined upper boundary and the less distinct lower boundary. They find

$$\frac{M_P}{M_J} \cong (1.7 \pm 0.2) \left(\frac{P}{day}\right)^{-1.14 \pm 0.14}$$

and

$$\frac{R_P}{R_J} \cong (1.4 \pm 0.3) \left(\frac{P}{day}\right)^{0.31 \pm 0.12}$$

for the upper boundary. WASP-12b itself was considered in the subset contributing to the derivation of the boundary in the ($M_P$, $P$) plane, but not in the ($R_P$, $P$) plane where it lay below the period range used. We can use the well-determined orbital period of WASP-12b to calculate the values of mass and radius predicted for a planet of this period which lies on the boundary of the sub-Jovian desert. Comparing these values with the less-precisely-determined empirical mass and radius for WASP-12b (see the introduction) we find that within the uncertainties, WASP-12b lies on both these boundaries, though it lies about 1σ on the low-mass, large radius side of the boundaries. Given that we see WASP-12b is currently losing mass, this is intriguing though inconclusive.

*Models to explain the sub-Jovian Desert*

There have been a number of mechanisms proposed to explain the observed sub-Jovian desert. The root cause(s) are not yet agreed, and it is possible that multiple mechanisms contribute to the explanation. Since WASP-12b lies at or near the well-defined upper boundary, we will focus predominantly on the explanations for this feature.

Mass loss due to the prodigious insolation suffered by short orbital period planets is an obvious candidate mechanism. Szabó & Kiss (2011) include this as the first of four possible hypotheses to explain their findings, in particular for planets with loosely gravitationally bound atmospheres. Indeed the correlation between surface gravity and orbital period found by Southworth et al (2007) and Southworth (2008) is in fact the first discovery of the upper boundary of the sub-Jovian desert.

Within the framework of planetary migration, Mazeh et al (2016) mention the possibility that the upper boundary acts as a death line: planets which migrate to shorter orbital period and cross this line lose the majority of their mass as a result of irradiation or Roche lobe-overflow. Upon crossing the death line, planets would move rapidly downwards in the ($M_P$, $P$) and ($R_P$, $P$) planes.

An alternative possibility sketched by Mazeh et al (2016) within the migration framework instead invokes a correlation between protoplanetary disk mass and planet mass. The disk pushes the planet inwards until the disk density diminishes below that required. More massive protoplanetary discs might persist to longer times and / or smaller distances from the star, thus growing more massive planets and pushing them further inwards.

Matsakos and Königl (2016) explain both the upper and lower boundaries simultaneously by considering the fate of planets which arrive at short orbital periods via eccentric orbits and subsequent circularisation. Planets can acquire an eccentric orbit with a small pericentre either via planet-planet scattering or more gradual processes such as Kozai-Lidov migration. If the pericentre of the eccentric orbit is less than the Roche limit $a_R$, the simplest form of which is

$$a_R = R_P \left(\frac{2M_*}{M_P}\right)^{\frac{1}{3}}$$

the planet will be tidally disrupted and will not survive to be subsequently discovered in a circular orbit. Consequently there is a minimum orbital radius after circularisation which depends on the planet's Roche limit. Since the Roche limit depends on the planet's mass and radius, this leads to different orbital criteria for surviving giant and terrestrial planets. Matsakos and Königl (2016) elegantly show that these criteria neatly lead to the observed upper and lower boundaries for the sub-Jovian desert.

Most recently Ginzburg and Sari (2017) treat the Roche lobe overflow of hot Jupiters analytically, and conclude that the sub-Jovian desert is a consequence of core masses > 6 $M_\oplus$ for the Galaxy's population of hot Jupiters.

*WASP-12b confronts the models*

To explain WASP-12b within Matsakos and Königl (2016)'s scenario, we need a precursor phase in which the planet acquired a high eccentricity low pericentre orbit. Since WASP-12 is the primary star of a hierarchical triple star system (Bechter et al 2014) this can naturally be explained by the secular dynamical evolution. Hamers and Lai (2017) present a simplified semi-analytic model of this evolution, concluding that extremely high eccentricities can be generated in the (WASP-12b) planetary orbit in this scenario. Note that the alternative planet-planet scattering route to a high eccentricity orbit for WASP-12b could be problematic because planet-planet scattering events are expected for only the first <$10^8$ years, at least ten times less than the age of WASP-12.

Kurokawa and Nakamoto (2014) consider the effects of XUV-driven atmospheric escape and Roche-lobe overflow on the exoplanet population. Their model assumes an insolation appropriate to a G type host star and WASP-12b rests comfortably above their minimum survival mass line for a planet with a $10 M_\oplus$ core. Clearly WASP-12b *is* losing mass, which can be reconciled with this model if the (unknown) XUV radiation flux on the planet exceeds that assumed and / or WASP-12b's mass loss rate is low enough that the planet will persist at essentially the same mass. Currently, as we have seen, the mass loss rate for WASP-12b is non-zero, but beyond that is very poorly constrained.

## Conclusions

WASP-12b is a fascinating planet, and has several extreme properties. It has a short orbital period, suffers very high insolation from its hot (F type) host star, and is among the largest known planets. The planet's large size is probably related to its proximity to the star, through irradiation and its highly filled Roche lobe. The optically opaque planet sits within the Roche lobe, but is surrounded by an exosphere of translucent gas which overfills the Roche lobe and produces line-blanketing absorption in the near-UV. The observed chromospheric emission from the host star WASP-12 is anomalously low: it is an extreme outlier in the distribution of the stellar chromospheric activity indicator log $R'_{HK}$. We attribute the anomalously depressed stellar chromospheric emission to absorption by a diffuse shroud of circumstellar gas which fills our line of sight, covering the chromospherically active regions of the star at all orbital phases. We surmise that the circumstellar gas shroud originates in mass loss from the planet, via the exosphere.

The mass loss rate from WASP-12b is unknown. Establishing a high precision measurement of the orbital eccentricity and monitoring the orbital period evolution over the coming years and decades are needed to predict and confront estimates of the mass loss rate. Observations which can achieve this include radial velocity measurements, transit timings, and secondary eclipse timings. An empirical determination of the ionising EUV flux experienced by WASP-12b is a further vital parameter. The best way to establish this is with a UV spectrum covering the C IV and Si IV features. Meanwhile the various ingredients needed to build computational models of the physical mechanisms underlying mass loss from an extreme planet like WASP-12b exist, but have not yet been fully integrated in a self-consistent way and applied to this case.

WASP-12b lies on the upper boundary of the sub-Jovian desert. Its position may be a natural consequence, through secular dynamical evolution, of the planet's status as the lowest mass object in a hierarchical quadruple system. It is unclear whether its current evolution will move it rapidly down in mass and radius through the desert; nor is the current and future orbital period evolution clear. Determination of the mass loss rate from WASP-12b will clarify the evolutionary pathway(s) leading to one of the most prominent demographic features of the Galaxy's population of exoplanets.

**Acknowledgements**
CH gratefully acknowledges financial support from STFC under grants ST/L000776/1 and ST/P000584/1, and thanks her many primary literature co-authors for sharing their knowledge, skills and enthusiasm. Special thanks to Joel and Charlotte for their forbearance while this Chapter was written.